# A Critical Analysis of Universality and Kirchhoff's Law: A Return to Stewart's Law of Thermal Emission

Pierre-Marie Robitaille, Ph.D. Dept. of Radiology, The Ohio State University,
130 Means Hall, 1630 Means Hall, 1654 Upham Drive, Columbus, Ohio 43210, USA
e-mail: robitaille.1@osu.edu

Abstract

It has been advanced, on experimental (P.-M. Robitaille, IEEE Trans. Plasma Sci. 2003, v. 31(6), 1263-1267) and theoretical (P.M. Robitaille, Progr. Phys. 2006, v.2, 22-23) grounds, that blackbody radiation is not universal and remains closely linked to the emission of graphite and soot. In order to strengthen such claims, a conceptual analysis of the proofs for universality is presented. This treatment reveals that Gustav Robert Kirchhoff has not properly considered the combined effects of absorption, reflection, and the directional nature of emission in real materials. In one instance, this leads to an unintended movement away from thermal equilibrium within cavities. Using equilibrium arguments, it is demonstrated that the radiation within perfectly reflecting or arbitrary cavities does not necessarily correspond to that emitted by a blackbody.

1 Introduction

Formulated in 1858, Stewart's Law [1] states that when an object is studied in thermal equilibrium, its absorption is equal to its emission [1]. Stewart's formulation leads to the realization that the emissive power of any object depends on its temperature, its nature, and on the frequency of observation. Conversely, Gustav Kirchhoff [2-4] reaches the conclusion that the emissive power of a body is equal to a universal function, dependent only on its temperature and the frequency of interest, and independent of its nature and that of the enclosure. He writes: *"When a space is surrounded by bodies of the same temperature, and no rays can penetrate through these bodies, every pencil in the interior of the space is so constituted, with respect to its quality and intensity, as if it proceeded from a perfectly black body of the same temperature, and is therefore independent of the nature and form of the bodies, and only determined by the temperature* (see [4], p. 96-97).*"*

At the same time, Max Planck, in his *Theory of Heat Radiation*, reminds us that: *"...in a vacuum bounded by totally reflecting walls any state of radiation may persist* (see [5], §51).*"* Planck is aware that a perfect reflector does not necessarily produce blackbody radiation in the absence of a perfect absorber [6]. It is not simply a matter of waiting a sufficient amount of time, but rather the radiation will *"persist"* in a non-blackbody, or arbitrary, state. Planck re-emphasizes this aspect when he writes: *"Every state of radiation brought about by such a process is perfectly stationary and can continue infinitely long, subject, however, to the condition that no trace of an emitting or absorbing substance exists in the radiation space. For otherwise, according to Sec. 51, the distribution of energy would, in the course of time, change through the releasing action of the substance irreversibly, i.e., with an increase of the total entropy, into the stable distribution corresponding to black radiation* (see [5], §91).*"* Planck suggests that if an absorbing substance is present, blackbody radiation is produced. Such a statement is not supported scientifically. In fact, a perfect absorber, such as graphite or soot, is required [6-8].

Recently, I have stated [6-8] that cavity radiation was not universal and could only assume the normal distribution (i.e. that of the blackbody) when either the walls of the cavity, or the objects it contains, were perfectly absorbing. These ideas are contrary to the expressed beliefs of Kirchhoff and Planck. Therefore, they deserve further exposition by revisiting Kirchhoff's basis for universality. In combination with a historical review of blackbody radiation [8], such an analysis demonstrates that claims of universality were never justified [6-8].

2.1 Kirchhoff's First Treatment of his Law

Kirchhoff's first presentation of his law [2] involved two plates, *C* and *c*, placed before one another (see Fig. 1). Neither plate was perfectly absorbing, or black. Behind each plate, there were mirrors, *R* and *r*, which ensured that all the radiation remained between the plates. Kirchhoff assumed that one of the plates, *c*, was made of a special material which absorbed only one wavelength and transmitted all others. This assumption appears to have formed the grounds for the most strenuous objections relative to Kirchhoff's first derivation [9-11]. Kirchhoff moved to insist (see [9] for a treatment in English) that, under these conditions, at a certain temperature and wavelength, all bodies had the same ratio of emissive and absorptive powers.

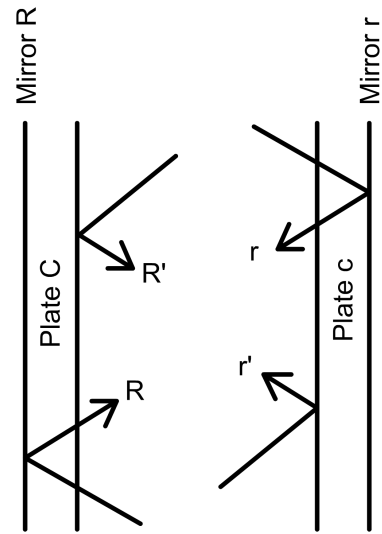

Figure 1. Schematic representation of Kirchhoff's first proof [2]. C and c represented objects of a specified nature (see text). *R* and *r* corresponded to perfectly reflecting mirrors. Note that Kirchhoff had neglected the reflection from the surfaces of *C* and *c* denoted as *R'* and *r'*.

The fallacy with Kirchhoff's argument lays not only in the need for a special material in the second plate, *c*, as so many have hinted [9-11]. The most serious error was that he did not consider the reflection from the plates themselves. He treated the reflection as coming only from the mirrors placed behind the plates. But this dealt with the problem of transmission, not reflection. As a result, Kirchhoff ignored the reflection produced by the surfaces of the plates.

The total radiation leaving from the surface of each plate, given thermal equilibrium, is obtained, not only by its emission, *E* (or *e*), but rather by the *sum* of its emission, *E (or e)*, and reflection, *R' (or r')*. It is only when the plates are black that surface reflection can be neglected. Consequently, if Kirchhoff insists that surface reflection itself need not be addressed ($R'=r'=0$), he simply proves that the ratio of emission to absorption is the same for all blackbodies, not for all bodies. The entire argument, therefore, is flawed because Kirchhoff ignored the surface reflection of each plate, and is considering all reflection as originating from the perfectly reflecting mirrors behind the plates. A proper treatment would not lead to universality, since the total radiation from plate *C* was *E + R'* not simply *E*, where *R'* denotes the reflection from surface *C* (see Fig.1). Similarly, the total radiation from plate *c* was *e + r'*, not simply *e*, where *r'* denotes the reflection from surface *c*. The mirrors, *R* and *r*, are actually dealing only with transmission through plates *C* and *c*. The conceptual difficulty when reviewing this work is that Kirchhoff apparently treats reflection, since mirrors are present. In fact, he dismisses the issue. The mirrors cannot treat the reflection off the surfaces of *C* and *c*. They deal with transmission. Kirchhoff's incorrect visualization of the effect of reflection is also a factor in his second proof.

2.2 Kirchhoff's Second Treatment of his Law

Kirchhoff's second treatment of his law [3, 4] is much more interesting conceptually and any error will consequently be more difficult to locate. The proof is complex, a reality recognized by Stewart in his *Reply*: *"I may remark, however, that the proof of the Heidelburg Professor is so very elaborate that I fear it has found few readers either in his own country or in this* [12].*"*

Kirchhoff began by imagining a cavity whose walls were perfectly absorbing (see Fig. 2). In the rear of the cavity was an enclosure wherein the objects of interest were placed. There were three openings in the cavity, labeled 1, 2, and 3. He conceived that openings 2 and 3 could each be sealed with a perfectly absorbing surface. As a result, when Kirchhoff did this, he placed his object in a perfectly absorbing cavity [6]. He eventually stipulated that the experiment was independent of the nature of the walls, in which case the cavity could be viewed as perfectly reflecting [6]. Yet, as has been previously highlighted [6], the scenario with the perfectly reflecting cavity required, according to Planck, the introduction of a minute particle of carbon [5, 8]. Hence, I have argued that Kirchhoff's analysis was invalid on this basis alone [6]. By carefully considering Kirchhoff's theoretical constructs, the arguments against blackbody radiation, within a perfect reflecting enclosure, can now be made from a slightly different perspective.

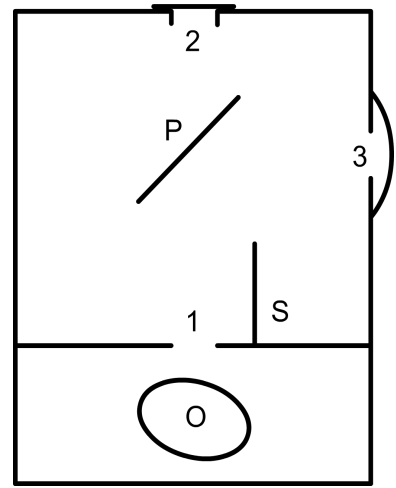

Figure 2. Schematic representation of Kirchhoff's second proof [3, 4]. The cavity contained three openings, labeled 1, 2, and 3. There was also a plate, *P*, which was perfectly transmitting for the frequency and polarization of interest, and perfectly reflecting for all others. While the existence of such a plate can be the source of objections relative to Kirchhoff's proof [10], the discussion in this work does not center on the nature of the plate. Idealized objects can be assumed as valid as they represent (more or less) mathematical extensions of physical observations (see text). A black screen, *S*, was used to prevent radiation from traveling directly between openings 1 and 3. An object, which was either perfectly absorbing or arbitrary, was placed in the enclosure located behind opening 1. The key to Kirchhoff's proof relied on rapidly changing the covering of opening 3, from a perfect concave mirror to a perfectly absorbing surface. In Kirchhoff's initial presentation, the entire cavity was perfectly absorbing [3, 4]. However, Kirchhoff extended his result to be independent of the nature of the walls, making it acceptable to consider the entire cavity as perfectly reflecting (see text).

Kirchhoff's analysis of his cavity (see Fig. 2) was ingenious. He set strict conditions for the positions of the walls which linked the openings 1 and 2, and which contained opening 3. The key was in the manner wherein opening 3 was handled. Kirchhoff permitted opening 3 to be covered either with a perfect absorber or with a perfect concave mirror. He then assumed that equilibrium existed in the cavity and that he could instantaneously change the covering at opening 3. Since equilibrium was always preserved, Kirchhoff could then treat the rays within the cavity under these two different conditions and, hence, infer the nature of the radiation within the cavity at equilibrium.

    Kirchhoff initially demonstrated that, if the enclosed object and the cavity were perfectly absorbing, the radiation was denoted by the universal function of blackbody radiation. He then replaced the object with an arbitrary one, and concluded, once again, that the radiation was black. Kirchhoff's presentation was elegant, at least when the cavity was perfectly absorbing. The Heidelburg Professor extended his findings to make them independent of the nature of the walls of the enclosure, stating that the derivation was valid, even if the walls were perfectly reflecting. He argued that the radiation within the cavity remained blackbody radiation. Let us revisit what Kirchhoff had done.

    Since the walls can be perfectly reflecting, this state is adopted for our analysis. Opening 3 can once again be covered, either by a concave mirror or by a perfectly absorbing surface. An arbitrary object, which is not a blackbody, is placed in the cavity. The experiment is initiated with the perfect concave mirror covering

opening 3. As shown in section 3.1.2, under these conditions, the cavity contains radiation whose nature depends not on the cavity, but on the object. This radiation, in fact, is not black. This can be seen, if the object was taken as perfectly reflecting. The arbitrary radiation is weaker at all frequencies. Thus, when an arbitrary object is placed in the enclosure, the intensity of the radiation within the cavity, at any given frequency, does not correspond to that predicted by the Planckian function (see section 3.1.2). However, when opening 3 is covered by a perfectly absorbing substance, the radiation in the cavity becomes black (see sections 3.1.2 and 3.2). The emission from the object is that which the object emits and which it reflects. The latter originates from the surface of opening 3 (see section 3.2). When the perfect absorber is placed over opening 3, the entire cavity appears to hold blackbody radiation. Therefore, by extending his treatment to the perfect reflector, Kirchhoff is inadvertently jumping from one form of cavity radiation (case 1: the concave mirror, object radiation) to another (case 2: the perfect absorber, blackbody radiation) when the covering on opening 3 is changed. At that moment, the cavity moves out of equilibrium.

Thus, Kirchhoff's proof is invalid. This is provided, of course, that the test began with the perfect concave mirror covering opening 3. Only under these circumstances would Kirchhoff's proof fail. Nonetheless, the experimental proof cannot be subject to the order in which manipulations are executed. This is because the validity of equilibrium arguments is being tested. Consequently, nothing is independent of the nature of the walls. This is the lesson provided to us by Balfour Stewart in his treatise when he analyzes radiation in a cavity temporarily brought into contact with another cavity [8]. Dynamic changes, not equilibrium, can be produced in cavities, if reflectors are used. This is the central error relative to Kirchhoff's second attempt at universality [3, 4].

There are additional minor problems in Kirchhoff's presentation [3, 4]. In §13 of his proof [3-4], Kirchhoff is examining an arbitrary object within a perfectly absorbing cavity. It is true that the resultant cavity radiation will correspond to a blackbody, precisely because the walls are perfectly absorbing (see section 3.1.1). However, Kirchhoff states: *"the law §3 is proved under the assumption that, of the pencil which falls from surface 2 through opening 1 upon the body C, no finite part is reflected by this back to the surface 2; further, that the law holds without limitation, if we consider that when the condition is not fulfilled, it is only necessary to turn the body C infinitely little in order to satisfy it, and that by such a rotation the quantities E and A undergo only and [sic] infinitely small change* (see [4], p.92)." Of course, real bodies can have diffuse reflection. In addition, rotation does not ensure that reflection back to surface 2 will not take place. Real bodies also have directional spectral emission, such that the effect of rotation on E and A is not necessarily negligible. These complications are of little significance within a perfectly absorbing cavity. The radiation within such enclosures is always black (see section 3.1.1). Conversely, the problems cannot be dismissed in the perfect reflector and the entire proof for universality, once again, is invalid.

For much of the 19th century, the understanding of blackbody radiation changed little, even to the time of Planck [11]. No laboratory proof of Kirchhoff's Law was ever produced, precisely because universality could not hold. Only theoretical arguments prevailed [10]. Yet, such findings cannot form the basis for a law of physics. Laws stem from experiments and are fortified by theory. They are not born *de novo*, using mathematics without further validation. It is not possible to ensure that black radiation exists, within a perfectly reflecting cavity, without recourse at least to a carbon particle [6, 8]. In fact, this is the route which Planck utilized in treating Kirchhoff's Law [5, 8].

3 Thermal Equilibrium in Cavities

A simple mathematical treatment of radiation, under conditions of thermal equilibrium, begins by examining the fate of the total incoming radiation, $\Gamma$, which strikes the surface of an object. The various portions of this radiation are either absorbed (A), reflected (R), or transmitted (T) by the object. If normalized, the sum of the absorbed, reflected, or transmitted radiation is equal to $\alpha + \rho + \tau = 1$. Here, absorptivity, $\alpha$, corresponds to the absorbed part of the incoming radiation/total incoming radiation. Similarly, the reflectivity, $\rho$, is the reflected part of the incoming radiation/total incoming radiation. Finally, the transmissivity, $\tau$, involves the transmitted part of the incoming radiation/total incoming radiation. If all objects under consideration are fully opaque, then $1 = \alpha + \rho$.

Stewart's Law [1] states that, under conditions of thermal equilibrium, the ability of an object to absorb light, α, is exactly equal to its ability to emit light, ε. Nonetheless, for this presentation, Stewart's Law is not assumed to be valid [1]. The question arises only in the final section (4.2), when two objects are placed within a perfectly reflecting cavity. Emissivity, ε, is standardized relative to lamp-black [8] and, for such a blackbody, it is equal to 1. For a perfect reflector, the emissivity, ε, is 0. All other objects hold values of emissivity between these two extremes. If thermal equilibrium is not established, then ε and α are not necessarily equal [8].

If a cubical cavity is considered with walls $P^1$, $P^2$, $P^3$, $P^4$, $P^5$ (top surface), and $P^6$ (bottom surface), the following can be concluded at thermal equilibrium: since $P^1$ and $P^3$ are equal in area and opposite one another, then the total radiation from these walls must be balanced, $\Gamma_{p1} - \Gamma_{p3} = 0$. Similarly, $\Gamma_{p2} - \Gamma_{p4} = 0$ and $\Gamma_{p5} - \Gamma_{p6} = 0$. As such, $\Gamma_{p1} = \Gamma_{p3}$ and $\Gamma_{p2} = \Gamma_{p4}$. If one considers pairs of adjacent walls, then $(\Gamma_{p1} + \Gamma_{p2}) - (\Gamma_{p3} + \Gamma_{p4}) = 0$. It is possible to conclude that $\Gamma_{p1} = \Gamma_{p2} = \Gamma_{p3} = \Gamma_{p4}$ and, using symmetry, it can finally be concluded that $\Gamma_{p1} = \Gamma_{p2} = \Gamma_{p3} = \Gamma_{p4} = \Gamma_{p5} = \Gamma_{p6}$. Consequently, with normalization, $\Gamma_c = (\Gamma_{p1} + \Gamma_{p2} + \Gamma_{p3} + \Gamma_{p4} + \Gamma_{p5} + \Gamma_{p6})/6$. For an opaque cavity, the total radiation coming from the cavity, $\Gamma_T$, is given by $\Gamma_T = \varepsilon_c \Gamma_c + \rho_c \Gamma_c = \varepsilon_c \Gamma_c + (1 - \alpha_c) \Gamma_c$. This states that the total emission from the cavity must be represented by the sum of its internal emission and reflection. If the cavity is constructed from perfectly absorbing walls, $\alpha_c = 1$, $\rho_c = 0$, yielding $\Gamma_T = \varepsilon_c \Gamma_c$. The cavity is black and $\varepsilon_c$ must now equal 1, by necessity. Stewart's Law [1] has now been proved for blackbodies. If the cavity is made from perfectly reflecting walls, at thermal equilibrium, $\varepsilon_c \Gamma_c + (1 - \alpha_c) \Gamma_c = 0$. There is also no source of radiation inside the cavity ($\varepsilon_c = 0$) and $(1 - \alpha_c) \Gamma_c = 0$, leading explicitly to $\Gamma_c = 0$. Because $\Gamma_c = 0$, the total radiation monitored $\Gamma_T = \varepsilon_c \Gamma_c + \rho_c \Gamma_c = 0$.

These conclusions can be extended to perfectly absorbing and reflecting cavities of rectangular (or arbitrary) shapes. The central point is that a perfectly reflecting cavity can sustain no radiation, a first hint that universality cannot be valid. Planck only obtains blackbody radiation, in such cavities, by invoking the action of a carbon particle [6, 8]. This special case will be treated in sections 3.1.1 and 3.2.

3.1 An Object in a Perfect Cavity

At thermal equilibrium, the total emission from the surface of the object, $\Gamma_{so}$, is equal to that from the surface of the cavity, $\Gamma_{sc}$. When normalizing, the total emission, $\Gamma_T$, will therefore be as follows: $\Gamma_T = \tfrac{1}{2} \Gamma_{so} + \tfrac{1}{2} \Gamma_{sc}$. The total radiation from the surface of the object is equal to that which it emits plus that which it reflects, $\Gamma_{so} = [\varepsilon_o \Gamma_o + \rho_o \Gamma_c]$, and similarly for the surface of the cavity, $\Gamma_{sc} = [\varepsilon_c \Gamma_c + \rho_c \Gamma_o]$. Therefore, at equilibrium, $[\varepsilon_o \Gamma_o + \rho_o \Gamma_c] = [\varepsilon_c \Gamma_c + \rho_c \Gamma_o]$ or $\Gamma_o [\varepsilon_o - \rho_c] = \Gamma_c [\varepsilon_c - \rho_o]$. Solving for either $\Gamma_o$ or $\Gamma_c$, we obtain that $\Gamma_o = \Gamma_c [\varepsilon_c - \rho_o]/[\varepsilon_o - \rho_c]$ and $\Gamma_c = \Gamma_o [\varepsilon_o - \rho_c]/[\varepsilon_c - \rho_o]$.

3.1.1 An Arbitrary Object in a Perfectly Absorbing Cavity ($\varepsilon_c = 1$, $\rho_c = 0$)

Since $\Gamma_T = \tfrac{1}{2} \Gamma_{so} + \tfrac{1}{2} \Gamma_{sc}$, then $\Gamma_T = \tfrac{1}{2} (\varepsilon_o \Gamma_c [\varepsilon_c - \rho_o]/[\varepsilon_o - \rho_c] + \rho_o \Gamma_c) + \tfrac{1}{2} (\varepsilon_c \Gamma_c + \rho_c \Gamma_c [\varepsilon_c - \rho_o]/[\varepsilon_o - \rho_c])$. It is readily shown that $\Gamma_T = \Gamma_c$. Note that no use of Stewart's Law [1] was made in this derivation. In any case, when an object is placed within a cavity, which is perfectly absorbing, the emitted spectrum is independent of the object and depends only on the nature of the cavity. A blackbody spectrum is produced. This was the condition which prevailed over much of the 19th century when cavities were often lined with soot [8]. If the radiation was *independent of the nature of the walls, or of the object*, it was because the walls were coated with this material [8].

3.1.2 An Arbitrary Object in a Perfectly Reflecting Cavity ($\varepsilon_c = 0$, $\rho_c = 1$)

Since $\Gamma_T = \tfrac{1}{2} \Gamma_{so} + \tfrac{1}{2} \Gamma_{sc}$, then $\Gamma_T = \tfrac{1}{2} (\varepsilon_o \Gamma_o + \rho_o \Gamma_o [\varepsilon_o - \rho_c]/[\varepsilon_c - \rho_o]) + \tfrac{1}{2} (\varepsilon_c \Gamma_o [\varepsilon_o - \rho_c]/[\varepsilon_c - \rho_o] + \rho_c \Gamma_o)$. It is readily shown that $\Gamma_T = \Gamma_o$. Note, once again, that no use of Stewart's Law [1] was made in this derivation. When an object is placed within a cavity which is perfectly reflecting, the emitted spectrum is determined only by the object and is independent of the nature of the cavity. If the object is perfectly absorbing, like a carbon particle [6, 8], a blackbody spectrum will be obtained. Furthermore, if an arbitrary object is placed within a cavity, which is perfectly reflecting, the emitted spectrum *is dependent only on the nature of the object*. One observes *object radiation*, not blackbody radiation, because the object was never black *a priori*. This is the condition which Kirchhoff has failed to realize when he extended his treatment to be independent of the nature of the walls in his 1860 proof [3, 4], as seen in section 2.

### 3.1.3 An Arbitrary Object in an Arbitrary Cavity

Since $\Gamma_T = \frac{1}{2} \Gamma_{so} + \frac{1}{2} \Gamma_{sc}$, then $\Gamma_T = \frac{1}{2} (\varepsilon_o \Gamma_o + \rho_o\Gamma_o [\varepsilon_o - \rho_c]/[\varepsilon_c - \rho_o]) + \frac{1}{2} (\varepsilon_c \Gamma_o [\varepsilon_o - \rho_c]/[\varepsilon_c - \rho_o] + \rho_c \Gamma_o ]$ or alternatively, $\Gamma_T = \frac{1}{2} (\varepsilon_o \Gamma_c [\varepsilon_c - \rho_o]/ [\varepsilon_o - \rho_c] + \rho_o\Gamma_c) + \frac{1}{2} (\varepsilon_c \Gamma_c + \rho_c\Gamma_c [\varepsilon_c - \rho_o]/[\varepsilon_o - \rho_c])$. In this case, the expressions cannot be further simplified and the initial form, $\Gamma_T = \frac{1}{2} \Gamma_{so} + \frac{1}{2} \Gamma_{sc}$, can be maintained. Therefore, the total radiation emitted from such a cavity is a mixture *depending on both the characteristics of the object and the walls of the cavity*. This highlights that cavities do not always contain black radiation and that universality is invalid [6-8].

### 3.2 An Arbitrary Object and a Carbon Particle in a Perfectly Reflecting Cavity

If thermal equilibrium exists between an opaque object, o, a carbon particle, p, and a cavity, c, then $[\varepsilon_o \Gamma_o + \rho_o\Gamma_p + \rho_o \Gamma_c] - [\varepsilon_p \Gamma_p + \rho_p \Gamma_o + \rho_p \Gamma_c] + [\varepsilon_c \Gamma_c + \rho_c \Gamma_o - \rho_c\Gamma_p] = 0$. Since the cavity is perfectly reflecting, $\Gamma_c=0$, $\varepsilon_c = 0$, and $\rho_c= 1$, yielding, $\varepsilon_o \Gamma_o + \rho_o\Gamma_P - \varepsilon_p \Gamma_p - \rho_p \Gamma_o + \Gamma_o - \Gamma_p = 0$, and with rearrangement, $(\varepsilon_o + \rho_o -1) \Gamma_p - \varepsilon_p \Gamma_p + (1-\rho_p) \Gamma_o = 0$. If we take Stewart's Law ($\varepsilon_p = \alpha_p$ ; $\varepsilon_o = \alpha_o$) as valid [1], we can see that $\varepsilon_o + \rho_o = 1$, and then $(1-\rho_p) \Gamma_o = \varepsilon_p \Gamma_p$, leading directly to $\Gamma_o = \Gamma_p$. Alternatively, we may notice that, by definition, $\rho_o = 1-\alpha_o$ and $\rho_p = 1-\alpha_p$, then, $\Gamma_o = [(\varepsilon_p - \varepsilon_o + \alpha_o)/ \alpha_p] \Gamma_p$. If we take the particle to be black, we can simplify to $\Gamma_o = (1 - \varepsilon_o + \alpha_o)\Gamma_p$. Therefore, if we then observe the radiation in the cavity and find it to be black, since the particle is also black, Stewart's law is verified. This is because $\Gamma_o$ will be black and equal to $\Gamma_p$ only when $\varepsilon_o = \alpha_o$.

The problem can be examined from a slightly different angle in order to yield a little more insight, but the same conclusions hold. Because the objects are in a perfect reflector, then the radiation coming off their surfaces can be expressed as $\Gamma_{so} = \varepsilon_o \Gamma_o + \rho_o\Gamma_p$ and $\Gamma_{sp} = \varepsilon_p \Gamma_p + \rho_p\Gamma_o$. Given thermal equilibrium, the production of radiation from each object must be equal, $\Gamma_{so} = \Gamma_{sp}$, and thus $\varepsilon_o \Gamma_o + \rho_o\Gamma_p = \varepsilon_p \Gamma_p + \rho_p \Gamma_o$. Consequently, $\Gamma_o = ([\varepsilon_p - \rho_o]/[\varepsilon_o - \rho_p])\Gamma_p$ (see section 3.1). If the particle is black, $\varepsilon_p =1$ and $\rho_p = 0$, and $\Gamma_o (1- \rho_o/ \varepsilon_o) = \Gamma_p$. As a result of thermal equilibrium, the object must be producing a total emission which appears black in nature. $\Gamma_o$ must equal $\Gamma_p$. All solutions involve $\rho_o + \varepsilon_o =1$, which as stated above, is a proof of Stewart's Law ($\varepsilon_o = \alpha_o$). The object takes the appearance of a blackbody through the sum of its emission and reflection. The presence of completely black radiation within a cavity filled in this manner constitutes an explicit verification of Stewart's Law [1], as mentioned above. Since such cavities are known to be black, Stewart's Law has been proven. In fact, we have returned to the first portion of section 3.1.2. The effect is the same as if the walls of the cavity were perfectly absorbing. This is the point Planck failed to realize when he placed the carbon particle within the perfectly reflecting cavity and gave it a catalytic function [5, 6, 8].

### 4 Conclusions

Nearly 150 years have now passed since Gustav Robert Kirchhoff first advanced his Law of Thermal Radiation. Kirchhoff's Law [2-4] was far reaching. Its universal nature had a profound effect on the scientists of the period. At the time, many of these men were trying to discover the most general laws of nature. Hence, the concept of universality had great appeal and became ingrained in the physics literature. As a result, Kirchhoff's Law has endured, despite controversy [10], until this day. Recently, I have questioned universality [6, 7]. It is doubtful that Kirchhoff's Law can long survive the careful discernment of those physicists who wish to further pursue this issue.

At the same time, Kirchhoff's Law seems inseparably tied to Max Planck's equation [13]. As such, could a reevaluation of Kirchhoff's ideas compromise those of Max Planck [13]? In the end, it is clear that this cannot be the case [8]. Planck's solution to the blackbody problem remains valid for cavities which are perfectly absorbing. Thus, physics loses nothing of the Planck and Boltzmann constants, *h* and *k*, which were born from the study of heat radiation [1, 8]. That blackbody radiation loses universal significance also changes nothing, in fact, relative to the mathematical foundations of quantum theory. However, the same cannot be said relative to experimental findings [8]. In the end, the physics community may well be led to reconsider some of these positions [8].

Balfour Stewart [1] preceded Kirchhoff [2-4] by more than two years in demonstrating, under equilibrium, the equality between absorptivity and emissivity. Stewart's treatment, unlike Kirchhoff's, does not lead to universality [1, 8, 9, 14] but, rather, shows that the emissive power of an object is dependent on its nature, its temperature, and the frequency of observation. This is true even within cavities, provided that they do not contain a perfect absorber. It is only in this special circumstance that the nature of the object is eliminated from the problem. Yet, this is only because the nature of the carbon itself controls the situation. Stewart also properly treats emission and reflection in his *Treatise* [14]. Despite popular belief to the contrary [9], Stewart's interpretation is the correct solution. Conversely, Kirchhoff's formulation, not only introduced error, but provided justification for setting temperatures inappropriately. I have repeatedly expressed concern in this area [6-8]. It can be argued that Stewart's analysis lacked mathematical sophistication [9]. Stewart himself [12] counters the point [8]. Nonetheless, it is doubtful that the important consequences of Stewart's work can continue to be ignored. Justice and the proper treatment of experimental data demand otherwise.

Acknowledgements: The author would like to thank Luc Robitaille for assistance in figure preparation.

Dedication: This work is dedicated to the memory of my beloved mother, Jacqueline Alice Roy. (May 12, 1935 – December 2, 1996).

References

1. Stewart B. An Account of Some Experiments on Radiant Heat, Involving an Extension of Prevost's Theory of Exchanges. *Trans. Royal Soc. Edinburgh.* 1858, v. 22(1), 1-20. (also found in *Harper's Scientific Memoirs* (J.S. Ames, Ed.) The Laws of Radiation and Absorption: Memoirs of Prévost, Stewart, Kirchhoff, and Kirchhoff and Bunsen (translated and edited by D.B. Brace), American Book Company, New York, 1901, pp. 21-50).
2. Kirchhoff G. Über den Zusammenhang zwischen Emission und Absorption von Licht und. Wärme. *Monatsberichte der Akademie der Wissenschaften zu Berlin, sessions of Dec.* 1859, 1860, 783-787.
3. Kirchhoff G. Über das Verhältnis zwischen dem Emissionsvermögen und dem Absorptionsvermögen. der Körper für Wärme und Licht. *Poggendorfs Annalen der Physik und Chemie*, 1860, v. 109, 275-301. (English translation by F. Guthrie - Kirchhoff G. On the Relation between the Radiating and the Absorbing Powers of Different Bodies for Light and Heat. *Phil. Mag.*, 1860, 4th Series, v. 20, pp. 1-21).
4. Kirchhoff G.. On the Relation Between the Emissive and the Absorptive Power of Bodies for Light and Heat. (Reprinted from "Investigations of the Solar Spectrum and the Spectra of the Chemical Elements, 2nd Edition, Berlin, Ferd. Dummler's Publishing House, 1866, Gesammelte Abhandlungen, pp. 571-598, Liepzig, 1882 as found in *Harper's Scientific Memoirs* (J.S. Ames, Ed.) The Laws of Radiation and Absorption: Memoirs of Prévost, Stewart, Kirchhoff, and Kirchhoff and Bunsen (translated and edited by D.B. Brace), American book company, New York, 1901, pp. 73-97).
5. Planck M. The Theory of Heat Radiation. P. Blakiston's Son& Co., Philadelphia, PA, 1914.
6. Robitaille P.M.L. On the Validity of Kirchhoff's Law of Thermal Emission. *IEEE Trans. Plasma Sci.*, 2003, v. 31(6), 1263-1267.
7. Robitaille P.M.L. An Analysis of Universality in Blackbody Radiation, *Prog. in Phys.*, 2006, v.2, 22-23.
8. Robitaille P.M.L. Blackbody Radiation and the Carbon Particle. *Prog. in Phys.*, 2008, v. 3, 36.
9. Siegel D.M. Balfour Stewart and Gustav Robert Kirchhoff: Two independent Approaches to Kirchhoff's Law. *Isis*, 1976, v. 67(4), 565-600.
10. Schirrmacher A. Experimenting Theory: The Proofs of Kirchhoff's Radiation Law before and After Planck. *Munich Center for the History of Science and Technology,* 2001. http://www.mzwtg.mwn.de/arbeitspapiere/Schirrmacher_2001_1.pdf
11. Cotton A. The Present Status of Kirchhoff's Law. *Astrophys. J.* , 1899, v. 9, 237-268.
12. Stewart B. Reply to Some Remarks by G. Kirchhoff in his Paper "On the History of Spectrum Analysis". *Phil. Mag.*, 1863, ser. 4, v. 25, 354-360.
13. Planck M. Über das Gesetz der Energieverteilung im Normalspektrum. *Annalen der Physik*, 1901, v. 4, 553-563.
14. Stewart B. An Elementary Treatise on Heat. Clarendon Press, Oxford, U.K., 1888. (available online: http://books.google.com).